\documentclass{appolb}
\usepackage{graphicx}

\usepackage{amsmath}
\usepackage{amssymb}
\usepackage{amsfonts}
\usepackage{graphicx}
\usepackage{pstricks}
\usepackage{tikz}
\usepackage{mathtools}

\begin{document}
\title{The Paper Title comes Here...%
\thanks{Presented at ...}%
}

\title{Bootstrap technique to study correlation between neutron skin thickness and the slope of symmetry energy in atomic nuclei}


\author{D. Muir$^a$,A. Pastore$^a$,  J. Dobaczewski$^{a-d}$ and C. Barton$^a$
\address{%
$^{a}$Department of Physics, University of York, Heslington, York, Y010 5DD, United Kingdom \\
$^{b}$Department of Physics, P.O. Box 35 (YFL), FI-40014 University of Jyv\"akyl\"a, Finland \\
$^{c}$Institute of Theoretical Physics, Faculty of Physics, University of Warsaw, ul. Pasteura 5, PL-02093 Warsaw, Poland \\
$^{d}$Helsinki Institute of Physics, P.O. Box 64, FI-00014 University of Helsinki, Finland\\
}}
\maketitle
\begin{abstract}
We present a new statistical tool based on random sampling to assess the confidence interval of Pearson's and Spearman's correlation coefficients. These estimators are then used to quantify the statistical correlations among the neutron skin thickness of atomic nuclei and the slope of the symmetry energy in the infinite nuclear medium. 
\end{abstract}
  
\section{Introduction}

The possible correlation observed between the neutron skin thickness in $^{208}$Pb and the slope of the symmetry energy in the infinite nuclear mediums has  attracted the interest of the scientific community~\cite{hor01}. The  symmetry energy and its derivative are crucial ingredients for the Equation of State of massive astrophysical objects such as neutron stars~\cite{lat07,gan12}.
Both quantities are strictly speaking pseudo-observables and they cannot be measured directly. They are typically extracted from other measurements on atomic nuclei~\cite{heb13} or more fundamental theoretical calculations~\cite{vid09,heb10}.

The neutron skin of heavy nuclei has been identified as a possible candidate to assess the slope of the symmetry energy in the nuclear medium $L_0$~\cite{cen09}. A series of theoretical studies have shown that using the suggested correlation among neutron skin thickness $\Delta r_{np}$ and $L_0$ it would be possible to extract reliable confidence intervals for $L_0$ by performing accurate measurements of $\Delta r_{np}$. This has been a strong motivation for an intense experimental campaign to measure the skin quantity with the best possible accuracy~\cite{abr12,tar14}.

In the present article, we perform an analysis of the suggested correlation in two doubly-magic tin isotopes. Compared to previous works, we discuss a new statistical tool named Bootstrap analysis. The latter provides us access to the simulated parent distribution of the estimator we apply on our data set and thus accessing the confidence interval without assuming a Gaussian error.

The article is organised as follows: in Sec.\ref{sec:skin} we revise the definition of neutron skin and in Sec.\ref{sec:boot} we introduce the Bootstrap method. Our results are discussed in Sec.\ref{sec:res} and in Sec.\ref{sec:conc} we present our conclusions.

\section{Neutron skin}\label{sec:skin}

The neutron skin thickness $\Delta r_{np}=\langle r^2_n\rangle^{1/2}-\langle r^2_p\rangle^{1/2}$ is defined as a difference between the radial extension of the neutron density against the proton one.
This quantity is extracted using a standard procedure based on the parametrisation of the nuclear density via a 2 parameter Fermi function (2pF)~\cite{war10} given by 

\begin{eqnarray}
\rho_q(r)=\frac{\rho_{0q}}{1+exp[(r-C_q)/a_q]}\;,
\end{eqnarray}

\noindent where $q=n,p$ is an index representing either a neutron (n) or a proton (p) respectively. $C_q,a_q$ are parameters fixed on the self-consistent nuclear density extracted from a fully self-consistent Hartree-Fock-Bogoliubov (HFB) calculation~\cite{war10,dob96}.
The values $\rho_{0q}$ are merely determined by the number of neutrons (N) and protons (Z) in a given nucleus.

Following Refs.~\cite{cen09,roc11}, we consider in our calculations different nuclear functionals, including both non-relativistic - Skyrme~\cite{sky59} and Gogny~\cite{dec80} functionals - and relativistic ones~\cite{rei89} as well. The Skyrme functionals have been selected from Ref.~\cite{dut12} to span as large a range of variation of the $L_0$ parameter as possible.
Moreover, we have also used different functionals with different density dependent forms~\cite{cha09} to enrich our model space.
 For the Gogny interaction we used D1S~\cite{dec80} and D1M~\cite{gor09b}. For the relativistic calculations we used the DD-ME Lagrangians~\cite{lal05}.

\section{Statistical tools}\label{sec:boot}

 Non-Parametric Bootstrap (NPB) is a statistical method introduced by Efron in 1979~\cite{efr79} as a statistical tool to evaluate the bias of some particular statistical estimators.
NPB is based on the simple assumption that any experimental data set contains information about its parent distribution thus, if our data set is sufficiently large, we can simply replace the original parent distribution via the empirical one obtained from the sample.
The latter is then approximated using a Monte Carlo method and performing a resampling of the original data set.
By exploring a large subset of the possible combinations (with repetitions) we can build out from our original data set and can then effectively build the empirical distribution for the given estimator.
%

The only hypothesis behind NPB is that the parent distribution has finite moments and the data of the sample are independent. 
In the present implementation, we use equal weighting for each data point, but more evolved bootstrap techniques based on Bayesian inference allow such an assumption to be bypassed~\cite{rub81}.

The use of NPB helps us in highlighting the possible presence of outlier points that may artificially drive the correlation. Since we allow repetitions, it is not guaranteed that in each re-sampled set we find \emph{all} the original data points. This means that the way correlations are calculated for each data set may differ. Having a peaked distribution would mean that the result is quite robust and independent of the particular choice of the data, a larger distribution would mean that a careful choice of the input data set may lead to different conclusions. In our case this would be reflected by a higher uncertainty on the resulting error bars of the estimator.

For the current article, we apply the NPB approach to the calculations of the Pearson ($r$) and Spearman ($\rho$) coefficients~\cite{hau11}. The first is a statistical test to verify a possible linear correlation amongst the data, while $\rho$  is a nonparametric measure of rank correlation. It assesses how well the data can be described using a monotonic function.
They are both constructed to vary from -1 to +1 with the same meaning: for $|r|=1$ ($|\rho|=1$) there is a clear linear (monotonic) correlation among the data and for $r=0$ ($\rho=0$) there is no such correlation.

We refer to Ref.~\cite{hau11} for an extensive discussion of the properties of the two estimators.
\section{Results}\label{sec:res}

\begin{figure}[htb]
\centerline{%
\includegraphics[width=0.45\textwidth,angle=-90]{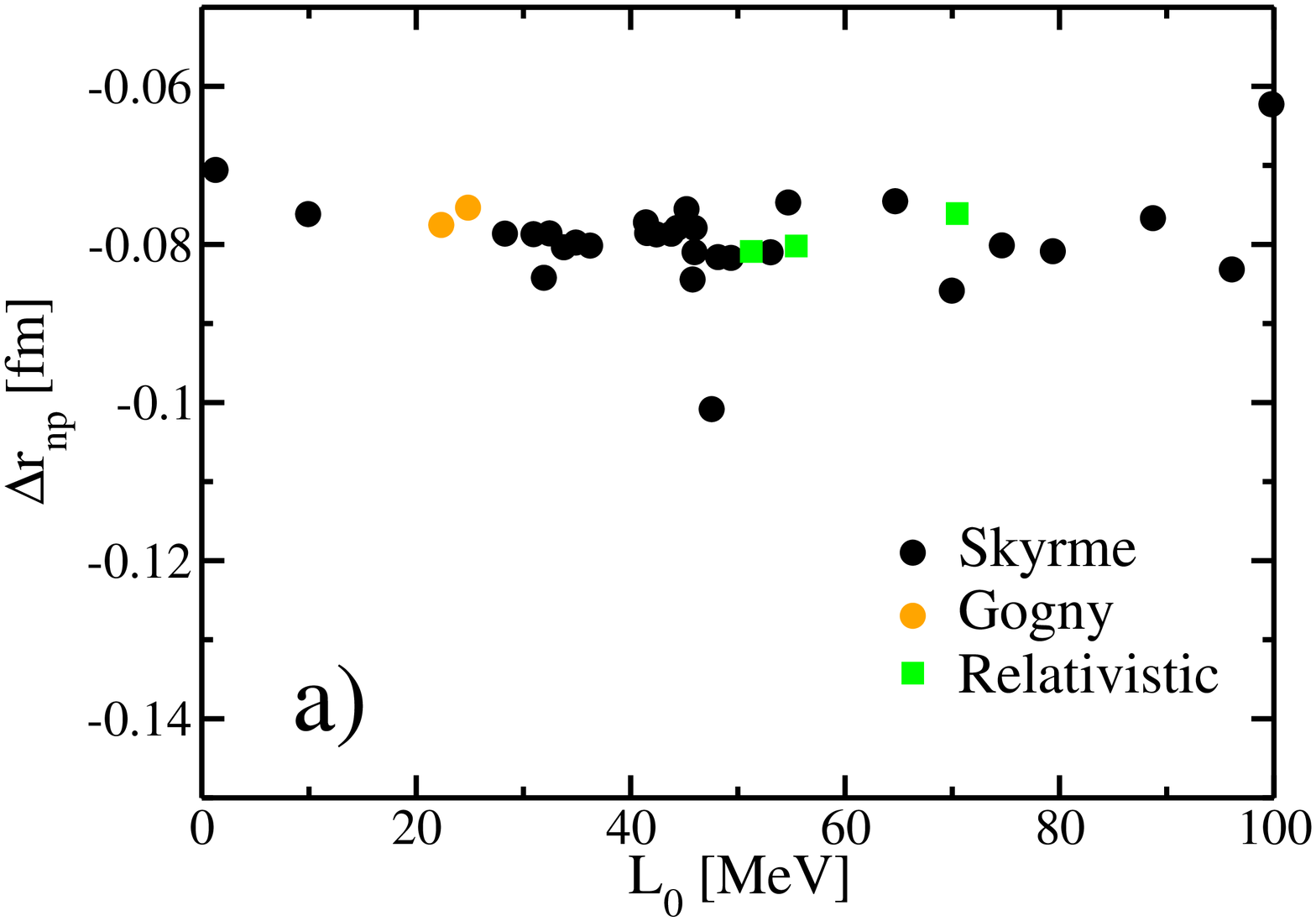}
\includegraphics[width=0.45\textwidth,angle=-90]{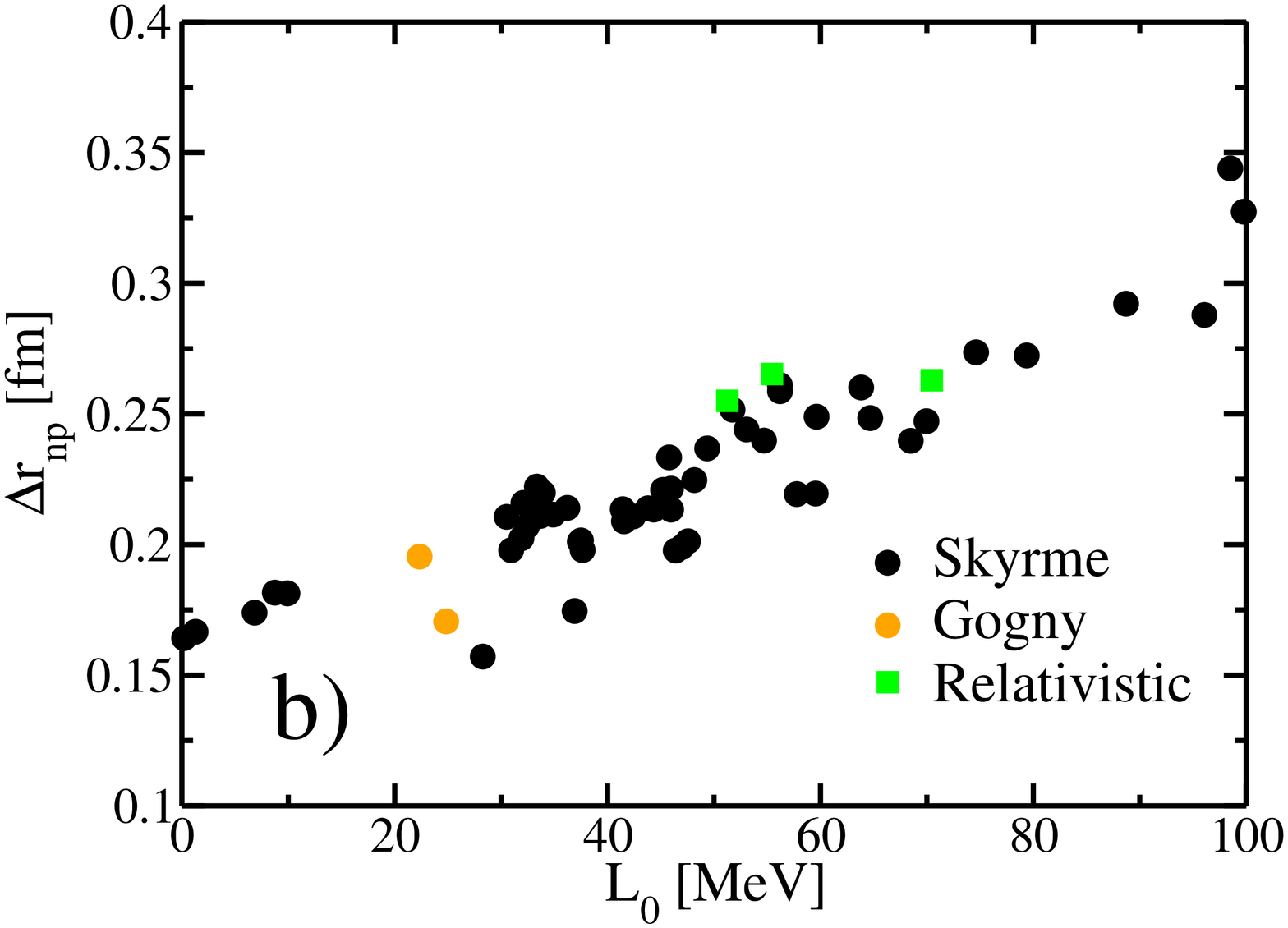}
}
\caption{(Colors online) Neutron skin thickness $\Delta r_{np}$ calculated for $^{100}$Sn (panel a) and $^{132}$Sn (panel b) as a function of $L_0$ for different functionals. See text for details.}
\label{Fig:sn100}
\end{figure}

In Fig.\ref{Fig:sn100} (a)-(b) we report the values of the neutron skin thickness in $^{100}$Sn (panel a) and $^{132}$Sn (panel b) as a function of the slope of the symmetry energy $L_0$. The latter is extracted as a density derivative of the symmetry energy parameter calculated at saturation density $\rho_{0}\backsim 0.16$ fm${}^{-3}$~\cite{dut12}.
We have used a total of 60 functionals spanning a very large interval of $L_0$. We observe that in the case of $^{100}$Sn, a small proton skin develops, mainly due to Coulomb repulsion since it is roughly independent of the particular choice of the $L_0$ parameter. In $^{132}$Sn a quite extended neutron skin develops and in this case we identify a clear trend as a function of $L_0$.

To quantify the possible correlation we apply both the Pearson and Spearman tests on these two sets of data. 
The results are reported in Tab.\ref{tab:values}.

\begin{table}
\begin{center}
\begin{tabular}{c|c}
\hline
\hline
\multicolumn{1}{c|}{$^{100}$Sn}&\multicolumn{1}{c}{$^{132}$Sn}\\
\hline
$r= 0.03_{-0.51}^{0.47}$ & $r=0.91_{-0.06}^{0.05}$\\[2mm]
$\rho=-0.17_{-0.36}^{0.38}$&$\rho=0.85_{-0.11}^{0.08}$  \\[2mm]
\hline
\hline
\end{tabular}
\end{center}
\caption{Pearson and Spearman coefficients for $^{100}$Sn and $^{132}$Sn. The errors refer to the 95\% confidence limits derived by inspecting the distribution of each estimator.}
\label{tab:values}
\end{table}

\begin{figure}[htb]
\centerline{%
\includegraphics[width=0.45\textwidth,angle=-90]{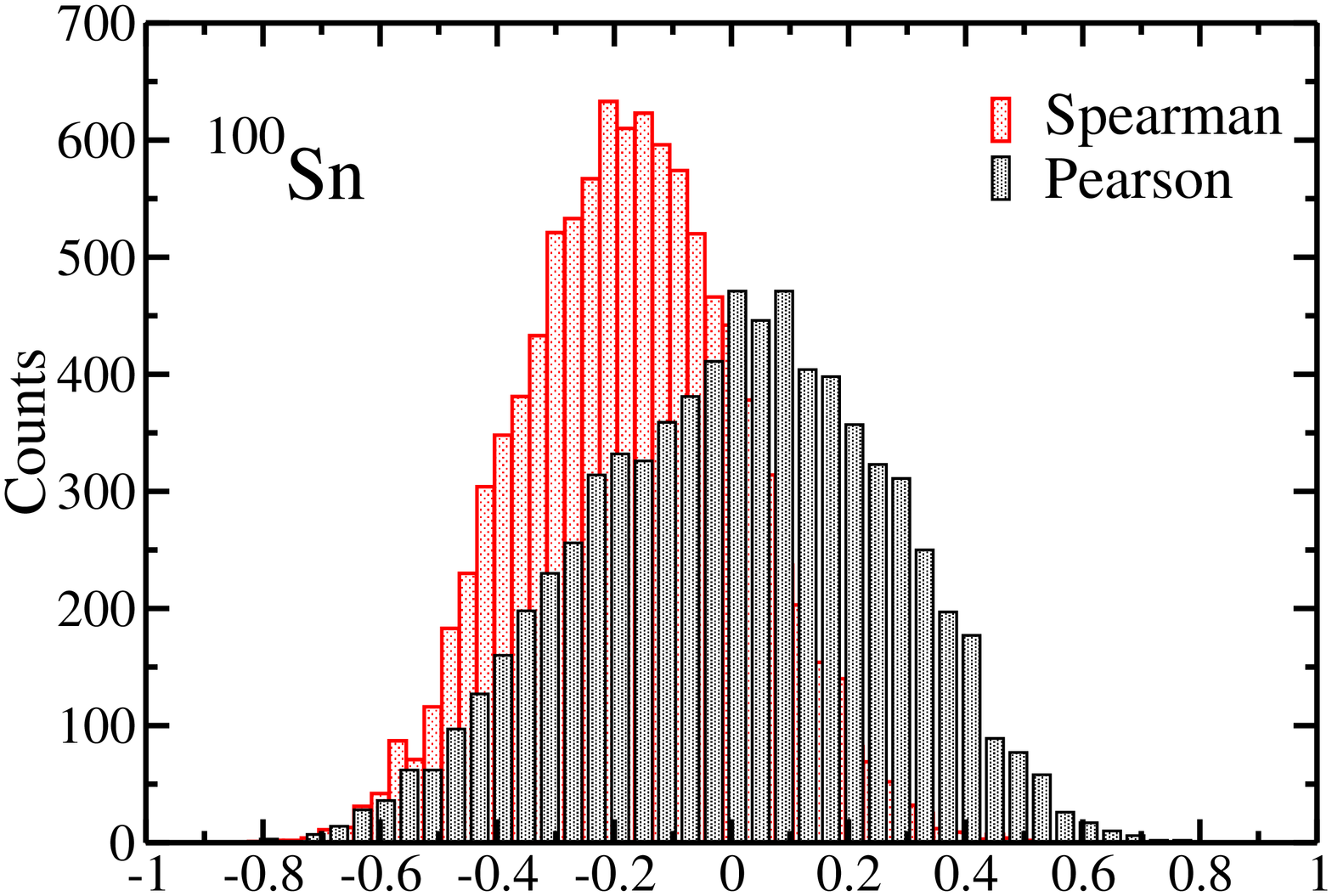}
\includegraphics[width=0.45\textwidth,angle=-90]{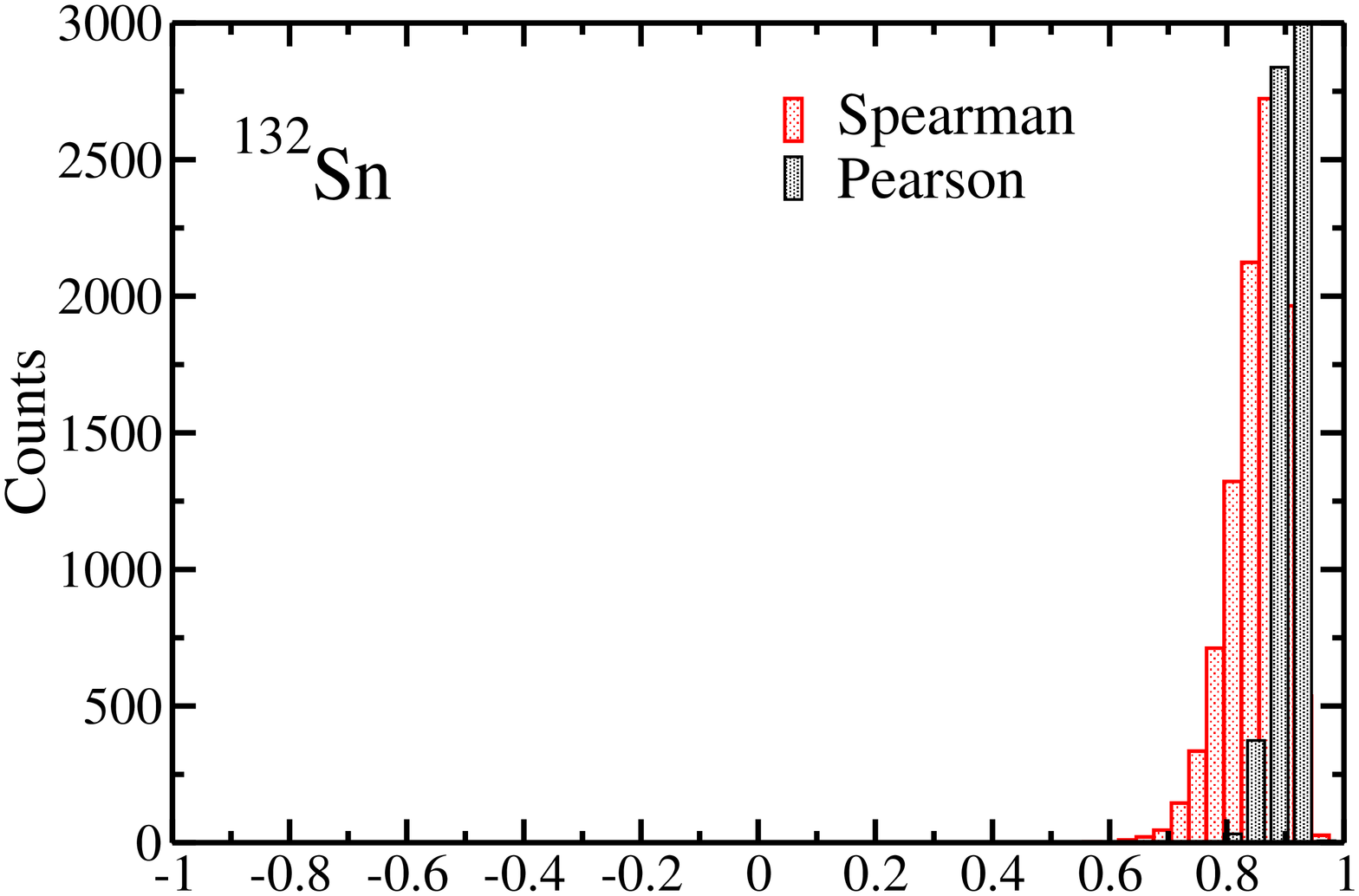}
}
\caption{(Colors online) Histograms of the $\hat{r}$ and $\hat{\rho}$ estimators as derived via the NPB approach for $^{100}$Sn (left panel) and $^{132}$Sn (right panel). See text for details.}
\label{Fig:histo}
\end{figure}

The error bars are defined as 95\% quantile of the distribution of the estimators for $\hat{r}$ and $\hat{\rho}$ extracted via the NPB and illustrated  in Fig.\ref{Fig:histo}. 
These histograms have been derived by performing 30,000 bootstrap simulations and performing for each of them the corresponding Pearson or Spearman correlation coefficient calculation.

We conclude that for $^{132}$Sn there is an observable linear correlation. For $^{100}$Sn there is no such correlation shown thus, the null-hypothesis of non-correlated data cannot be rejected. It is worth recalling that the statistical test evaluates the correlation amongst the data set but it does not establish if such a correlation is related to a physical phenomenon or it is an artefact of the model used to extract the data.

\section{Conclusions}\label{sec:conc}
We have applied a new statistical tool named Non-Parametric Bootstrap to assess the error bars of Pearson and Spearman correlation estimators. The main advantage of the NPB method is to produce robust confidence intervals that allow us to make the result on the estimator independent of the particular choice of the input data set used.

We have applied such a method to two doubly-magic nuclei and we have observed that we have a relatively high statistical correlation between the resulting neutron skin in $^{132}$Sn and the slope of the symmetry energy, while we have no significant statistical correlation between the proton skin of $^{100}$Sn and $L_0$. 

In the near future we plan to apply such a technique to a large variety of nuclei to assess the evolution of the statistical correlation as a function of the isospin asymmetry.
\section*{Acknowledgements}
 This work is supported  by the UK Science and Technology Facilities Council under Grants No. ST/L005727 and ST/M006433. 

\bibliographystyle{polonica}
\bibliography{biblio}

\end{document}